\documentclass[a4paper]{jpconf}
\usepackage{graphicx,amssymb}
\begin{document}
\begin{flushright}
 OU-HET-690
\end{flushright}

\vspace{-10mm}
\title{Gauge-invariant decomposition of nucleon spin}

\author{Masashi Wakamatsu}

\address{Department of Physics, Faculty of Science, Osaka University, 
Osaka 560-0043, JAPAN}

\ead{wakamatu@phys.sci.osaka-u.ac.jp}

\begin{abstract}
Based on gauge-invariant decomposition of covariant
angular momentum tensor of QCD in an arbitrary Lorentz frame,
we investigate the relation between the known decompositions of
the nucleon spin into its constituents, thereby clarifying in what
respect they are common and in what respect they are different
critically. We argue that the decomposition of Bashinsky and Jaffe,
that of Chen et al., and that of Jaffe and Manohar are contained
in our more general decomposition, after an appropriate
gauge-fixing in a suitable Lorentz frame, which means that they
all {\it gauge-equivalent}. We however point out that there is
another gauge-invariant decomposition of the nucleon spin,
which is closer to the Ji decomposition, while allowing the decomposition
of the gluon total angular momentum into its spin and orbital parts.
An advantage of the latter decomposition is that each of the four terms
corresponds to a definite observable, which can be extracted from
high-energy deep-inelastic-scattering measurements.
\end{abstract}

\section{Introduction}

After 20 years of theoretical and experimental efforts,
we now believe that only about 1/3 of the nucleon spin comes from the
intrinsic quark spin \cite{EMC88}\nocite{EMC89}
\nocite{COMPASS07}-\cite{HERMES07}.
Unfortunately, what carry the remaining 2/3 of
the nucleon spin is still a totally unanswerable question.
Especially difficult question here is whether the gluon total angular
momentum can be decomposed into its spin and orbital parts in a
gauge-invariant way.
Most people believe that the polarized gluon distribution $\Delta g(x)$
is an observable quantity from polarized DIS measurements. 
On the other hand, it is often claimed that there is no gauge-invariant
decomposition of gluon total angular momentum into its spin and orbital
parts.
Because the gauge principle is one of the most important principle of physics,
which demands that only gauge-invariants can be observed, how to reconcile
these conflicting observations is a fundamentally important problem in
nucleon spin physics.


\begin{figure}[h]
\begin{center}
\includegraphics[width=9.0cm]{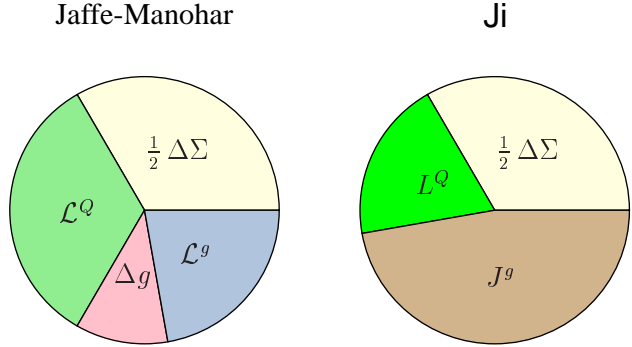}
\caption{\label{label}Two well-known nucleon spin decompositions.}
\end{center}
\end{figure}

First, we recall that there are two popular decompositions of the
nucleon spin. One is the Jaffe-Manohar decomposition \cite{JM90}
given in the form : 
\begin{eqnarray}
 \mbox{\boldmath $J$}_{QCD} &=&
 \int \psi^\dagger \,\frac{1}{2} \, 
 \mbox{\boldmath $\Sigma$} \,
 \psi \,d^3 x \ + \ 
 \int \psi^\dagger \,\mbox{\boldmath $x$} \times
 \frac{1}{i} \,
 \nabla \,\psi \,d^3 x \nonumber \\ 
 &+& \int \mbox{\boldmath $E$}^a \times
 \mbox{\boldmath $A$}^a \,d^3 x \ + \ 
 \int E^{a i} \,\mbox{\boldmath $x$} \times 
 \nabla \,A^{a i} \,d^3 x ,
\end{eqnarray}
while the other is the Ji decomposition \cite{Ji97} given as follows :
\begin{eqnarray}
 \mbox{\boldmath $J$}_{QCD} &=&
 \int \psi^\dagger \,\frac{1}{2} \,
 \mbox{\boldmath $\Sigma$} \,
 \psi \,d^3 x \ + \ 
 \int \psi^\dagger \,\mbox{\boldmath $x$} \times
 \frac{1}{i} \,
 \mbox{\boldmath $D$} \,
 \psi \,d^3 x \ + \ 
 \int \mbox{\boldmath $x$} \times
 (\mbox{\boldmath $E$}^a \times
 \mbox{\boldmath $B$}^a ) \,d^3 x .
\end{eqnarray}
In these widely-known decompositions, only the
intrinsic quark spin part is common, and the other parts are
totally different.
An apparent disadvantage of the Jaffe-Manohar decomposition is that
each term is not separately gauge-invariant,
except for the quark spin part. 
On the other hand, each term of the Ji decomposition is
separately gauge-invariant. Note, however, that further gauge-invariant
decomposition of $J^g$ into its spin and orbital parts is given up in
this widely-known decomposition.
An especially important observation here is that, since the quark orbital
angular momenta (OAMs) in the two decompositions are apparently different, i.e. 
\begin{equation}
 {\cal L}^Q \ \neq \ L^Q ,
\end{equation}
one must necessarily conclude
that the sum of the gluon spin and OAM in the Jaffe-Manohar decomposition
does not coincide with the total gluon spin in the Ji decomposition, i.e.
\begin{equation}
 \Delta g \ + \ {\cal L}^g \ \neq \ J^g .
\end{equation}

Recently, a new gauge-invariant decomposition of nucleon spin has been
proposed by Chen et al. \cite{Chen08},\cite{Chen09}.
The basic idea is to decompose the gluon field $\mbox{\boldmath $A$}$ into
two parts, the physical part $\mbox{\boldmath $A$}_{phys}$ and
the pure-gauge part $\mbox{\boldmath $A$}_{pure}$.
Imposing some additional condition, i.e. what-they-call the generalized
Coulomb gauge condition, Chen et al. arrive at the decomposition of the
nucleon spin in the following form : 
\begin{eqnarray}
 \mbox{\boldmath $J$}_{QCD} &=& \int \,\psi^\dagger \,
 \frac{1}{2} \,\mbox{\boldmath $\Sigma$} \,\psi \,d^3 x \ + \ 
 \int \psi^\dagger \,\mbox{\boldmath $x$} \times 
 ( \mbox{\boldmath $p$} - g \,\mbox{\boldmath $A$}_{pure} )
 \,\psi \,d^3 x \nonumber \\
 &+& \int \,\mbox{\boldmath $E$}^a \times 
 \mbox{\boldmath $A$}^a_{phys} \,d^3 x
 \ + \
 \int \,E^{aj} \,(\mbox{\boldmath $x$} \times \nabla ) \,
 \mbox{\boldmath $A$}^{aj}_{phys}  
 \,d^3 x \nonumber \\
 &=& \mbox{\boldmath $S$}^{\prime q} \ + \ 
 \mbox{\boldmath $L$}^{\prime q} \ + \ 
 \mbox{\boldmath $S$}^{\prime g} \ + \ 
 \mbox{\boldmath $L$}^{\prime g} .
\end{eqnarray}
An interesting feature of this decomposition is that each term is
separately gauge-invariant, while allowing the decomposition of the
gluon total angular momentum into its spin and orbital parts.
Another noteworthy feature of this decomposition is that it reduces to
the gauge-variant decomposition of Jaffe and Manohar in a particular
gauge, $\mbox{\boldmath $A$}_{pure} = 0, \mbox{\boldmath $A$} = 
\mbox{\boldmath $A$}_{phys}$.

In a recent paper \cite{Waka10A}, however, we have shown that the way
of gauge-invariant decomposition of nucleon spin is not necessarily unique,
and proposed yet another gauge-invariant decomposition given as follows :
\begin{eqnarray}
 \mbox{\boldmath $J$}_{QCD} &=& 
 \mbox{\boldmath $S$}^{q} \ + \ 
 \mbox{\boldmath $L$}^{q} \ + \ 
 \mbox{\boldmath $S$}^{g} \ + \ 
 \mbox{\boldmath $L$}^{g} ,
\end{eqnarray}
where
\begin{eqnarray}
 \mbox{\boldmath $S$}^q &=& \int \,\psi^\dagger \,
 \frac{1}{2} \,\mbox{\boldmath $\Sigma$} \,\psi \,d^3 x , \\
 \mbox{\boldmath $L$}^q &=& \int \,\psi \,
 \mbox{\boldmath $x$} \times
 (\mbox{\boldmath $p$} - g \,\mbox{\boldmath $A$} )
 \,\psi \,d^3 x , \\
 \mbox{\boldmath $S$}^g &=& \int \,
 \mbox{\boldmath $E$}^a \times 
 \mbox{\boldmath $A$}^a_{phys} \,d^3 x , \\
 \mbox{\boldmath $L$}^g &=& \int \,E^{aj} \,
 (\mbox{\boldmath $x$} \times \nabla) \,
 A^{aj}_{phys} \,d^3 x \ + \ 
 \int \,\rho^a \,(\mbox{\boldmath $x$} \times 
 \mbox{\boldmath $A$}^a_{phys} ) \,d^3 x . \label{Eq-Lg}
\end{eqnarray}
The characteristic features of our decomposition are as follows.
First, the quark parts of this decomposition is common with the Ji
decomposition. Second, the quark and gluon spin parts are common
with the Chen decomposition. A crucial difference with the Chen
decomposition appears in the orbital parts.
That is, although the sums of the quark and gluon OAMs in the two
decompositions are the same, i.e.
\begin{equation}
 \mbox{\boldmath $L$}^q \ + \ \mbox{\boldmath $L$}^g
 \ = \  
 \mbox{\boldmath $L$}^{\prime q} \ + \ 
 \mbox{\boldmath $L$}^{\prime g} ,
\end{equation}
each term is different in such a way that 
\begin{equation}
 \mbox{\boldmath $L$}^g - \mbox{\boldmath $L$}^{\prime g}
 \ = \ 
  - \,
 (\,\mbox{\boldmath $L$}^q - \mbox{\boldmath $L$}^{\prime q})
 \ = \ 
 \int \,\rho^a \,(\mbox{\boldmath $x$} \times 
 \mbox{\boldmath $A$}^a_{phys}) \,d^3 x .
\end{equation}
The difference arises from the treatment of the 2nd term of Eq.(\ref{Eq-Lg}).
We call this term the {\it potential angular momentum} term, since
the QED correspondent of this term is the orbital angular momentum
carried by the electromagnetic field or potential, which appears
in the Feynman paradox raised in his famous textbook of classical
electrodynamics \cite{BookFeynman}.
We have included this term into the {\it gluon} OAM part,
while Chen et al. included it into the {\it quark} OAM part.

To explain it in more detail, we first recall that that the potential angular momentum
term can also be expressed as
\begin{equation}
 \int \,\rho^a (\mbox{\boldmath $x$} \times \mbox{\boldmath $A$}^a_{phys})
 \,d^3 x \ = \ g \,\int \,\psi^\dagger \,
 \mbox{\boldmath $x$} \times \mbox{\boldmath $A$}_{phys} \,\psi \,d^3 x .
\end{equation}
Note that this term is solely gauge-invariant, as can easily be convinced from
the covariant (or homogeneous) gauge transformation property of the physical
part of the gluon field $\mbox{\boldmath $A$}_{phys}$.
This means that the gauge principle alone cannot say in which part of the
decomposition one should include the potential angular momentum term.
One has a freedom to include it into the quark OAM part, which would lead
to the Chen decomposition.
In fact, one sees that, in the following sum, the physical part of gluon field is
exactly canceled out and the pure gauge part remains, which just corresponds
to the quark OAM part of the Chen decomposition : 
\begin{eqnarray}
 &\,& \mbox{\boldmath $L$}^q \,(\mbox{Ours}) 
 \ + \ \mbox{\rm potential angular momentum}
 \nonumber \\
 &=&
 \int \,\psi^\dagger \,\mbox{\boldmath $x$} \times 
 (\mbox{\boldmath $p$} - g \,\mbox{\boldmath $A$} ) \,
 \psi \,d^3 x \ + \ g \,\int \,\psi^\dagger \,
 \mbox{\boldmath $x$} \times \mbox{\boldmath $A$}_{phys} \,
 \psi \,d^3 x \nonumber \\
 &=&
 \int \,\psi^\dagger \,\mbox{\boldmath $x$} \times 
 (\mbox{\boldmath $p$} - g \,\mbox{\boldmath $A$}_{pure} \,) \,
 \psi \,d^3 x \ \ \  = \ \ \ 
 \mbox{\boldmath $L$}^{\prime q} \,(\mbox{Chen}) .
\end{eqnarray}
However, we do not recommend the Chen decomposition, because
the common knowledge of standard electrodynamics tells us that the
momentum appearing in the Lorentz force equation of motion, i.e.
\begin{eqnarray}
 m \,\frac{d^2 \mbox{\boldmath $x$}}{d t^2} \ = \ 
 \frac{d \mbox{\boldmath $\Pi$}}{dt} \ = \ 
 e \,\left[\,\mbox{\boldmath $E$} \ + \ \frac{1}{2} \,
 \left(\,\frac{d \mbox{\boldmath $x$}}{d t} \times 
 \mbox{\boldmath $B$}
 \ - \ \mbox{\boldmath $B$} \times 
 \frac{d \mbox{\boldmath $x$}}{d t} \,
 \right) \,\right] .
\end{eqnarray}
is the so-called dynamical momentum
$\mbox{\boldmath ${\Pi}$} = \mbox{\boldmath $p$}
- e \,\mbox{\boldmath $A$}$
with the full gauge field $\mbox{\boldmath $A$}$,
not the canonical momentum $\mbox{\boldmath $p$}$ or 
its nontrivial gauge-invariant
extension $\mbox{\boldmath $p$} - e \,\mbox{\boldmath $A$}_{pure}$
\cite{BookSakurai}. By the same token,
the orbital angular momentum, accompanying the mass
flow of a charged particle, is the {\it dynamical} OAM,
not the {\it canonical} OAM.

Since the quark part of our decomposition just coincides with the Ji
decomposition, we may follow Ji's well-known program toward the full
decomposition of the nucleon spin.
First, extract $J^q$ and $J^g$ through GPD analyses.
Next, extract the quark and gluon polarization through polarized DIS
measurements and identify them with the quark and gluon spin parts
in our decomposition.
Then, the quark and gluon OAM can be known by subtracting $S^q$ and
$S^g$ from $J^q$ and $J^g$, respectively.

What was lacking in this wishful argument is a rigorous proof of the
identification of our gluon spin part with the 1st moment of the
polarized gluon distribution $\Delta g(x)$.
This problem is of fundamental importance, especially
because we are aware of the wide-spread statement that there is no
gauge-invariant decomposition of $J^g$, and that there is no
gauge-invariant local operator corresponding to the 1st moment of
$\Delta g(x)$.
Another important problem is as follows.
Since our gauge-invariant decomposition was given in a particular or fixed
Lorentz frame, we could not give a definite answer to the question whether
our decomposition has a frame-independent meaning or not?
The confirmation of frame-independence is very important. Otherwise,
the decomposition cannot be thought of as reflecting an intrinsic property
of the nucleon. 
We can show that these two questions can be answered simultaneously, by
making full use of a gauge-invariant decomposition of 
covariant angular-momentum tensor of QCD in an arbitrary Lorentz
frame \cite{Waka10B}.

\section{Gauge-invariant decomposition of covariant angular-momentum tensor}

Covariant generalization of the gauge-invariant decomposition of the
nucleon spin has twofold advantages.
Firstly, it is vital to prove the Lorentz frame-independence of the
decomposition.
Secondly, it can generalize and unify the various nucleon spin decompositions
in the market.
To achieve this goal, we can follow a similar idea as proposed by Chen et
al. \cite{Chen08},\cite{Chen09}.

The startingpoint is the decomposition
of the full gauge field $A^\mu$ into its physical and pure-gauge parts, i.e. 
$A^\mu = A^\mu_{phys} + A^\mu_{pure}$.
Here, we impose the following conditions on those components.
The first is the pure-gauge condition for $A^\mu_{pure}$ : 
\begin{eqnarray}
 F^{\mu \nu}_{pure} \ \equiv \ 
 \partial^\mu \,A^\nu_{pure} - 
 \partial^\nu \,A^\mu_{pure} - i \,g \,
 [A^\mu_{pure}, A^\nu_{pure}] 
 \ = \ 0 ,
\end{eqnarray}
while the second is the
gauge transformation properties for these two components :
\begin{eqnarray}
 A^\mu_{phys}(x) &\rightarrow&
 U(x) \,A^\mu_{phys}(x) \,U^{-1}(x) , \\
 A^\mu_{pure}(x) &\rightarrow&
 U(x) \,\left(\,A^\mu_{pure}(x) - \frac{i}{g}
 \,\, \partial^\mu \,\right) \,U^{-1}(x) .
\end{eqnarray}
As a matter of course, these conditions are not enough to fix gauge
uniquely. However, the point of our argument is that we can postpone
a {\it concrete} gauge fixing until later stage, while accomplishing a
gauge-invariant decomposition of $M^{\mu \nu \lambda}$ based on the
above conditions only.
As expected, we again find that the way of gauge-invariant decomposition
is not unique. Basically, we are left with two possibilities, i.e.
the decomposition (I) and the decomposition (II) \cite{Waka10B}.

The decomposition (II) is given as follows : 
\begin{eqnarray}
 M^{\mu \nu \lambda}_{QCD} &=& 
 M^{\prime \mu \nu \lambda}_{q-spin} \ + \ 
 M^{\prime \mu \nu \lambda}_{q-OAM} \ + \ 
 M^{\prime \mu \nu \lambda}_{g-spin} \ + \ 
 M^{\prime \mu \nu \lambda}_{g-OAM} \nonumber \\
 &+& \ \mbox{\rm boost} 
 \ + \ 
 \mbox{\rm total divergence} ,
\end{eqnarray}
with
\begin{eqnarray}
 M^{\prime \mu \nu \lambda}_{q-spin}
 &=& \frac{1}{2} \,\epsilon^{\mu \nu \lambda \sigma} \,
 \bar{\psi} \,\gamma_{\sigma} \,\gamma_5 \,\psi , \\
 M^{\prime \mu \nu \lambda}_{q-OAM} 
 &=& \bar{\psi} \,\gamma^{\mu} \,(\,x^{\nu} \,i \,
 D^{\lambda}_{pure}  
 \ - \ x^{\lambda} \,i \,
 D^{\nu}_{pure} \,) \,\psi , \\
 M^{\prime \mu \nu \lambda}_{g-spin}
 &=& 2 \,\mbox{Tr} \,\{\, F^{\mu \lambda} \,A_{phys}^{\nu} \,
 \ - \ F^{\mu \nu} \,A_{phys}^{\lambda} \,\} , \\
 M^{\prime \mu \nu \lambda}_{q-OAM}
 &=& 2 \,\mbox{Tr} \,\{\, F^{\mu \alpha} \,
 (\,x^{\nu} \,D_{pure}^{\lambda}
 \ - \ x^{\lambda} \,D_{pure}^{\nu} \,) \,A_{\alpha}^{phys} \,\} .
\end{eqnarray}
Here, the 1st and the 2nd terms are the quark spin and OAM terms, while 
the 3rd and the 4th terms are the gluon spin and OAM terms.
The remaining boost and the total divergence terms do not contribute to
the nucleon spin sum rule so that they can be neglected in the following
discussion.
Remarkably, we can show that this decomposition reduces to any ones of
Bashinsky-Jaffe \cite{BJ99}, of Chen et al. \cite{Chen08},\cite{Chen09},
and of Jaffe-Manohar \cite{JM90}, after an appropriate
gauge-fixing in a suitable Lorentz frame, which means that they are all
{\it gauge-equivalent} !
However, they are not our recommendable decompositions, because the quark
and gluon OAMs in those do not correspond to known experimental observables. 

\begin{figure}[h]
\begin{center}
\includegraphics[width=12cm]{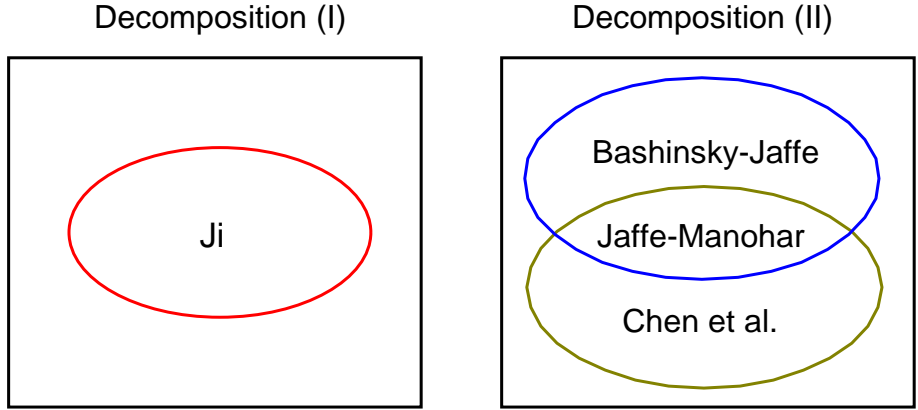}
\caption{\label{label}Schematic picture of two independent gauge-invariant
decompositions of nucleon spin and the relation with the known decompositions.}
\end{center} 
\end{figure}

Our recommendable decomposition is the gauge-invariant decomposition (I),
which is given in the following form : 
\begin{eqnarray}
 M^{\mu \nu \lambda} &=& M^{\mu \nu \lambda}_{q - spin}
 \ + \ M^{\mu \nu \lambda}_{q - OAM} \ + \ M^{\mu \nu \lambda}_{g - spin}
 \ + \ M^{\mu \nu \lambda}_{g - OAM} \nonumber \\
 &+& \ \mbox{\rm boost}  
 \ \ + \ \ \mbox{total divergence} ,
\end{eqnarray}
with
\begin{eqnarray}
 M^{\mu \nu \lambda}_{q - spin}
 &=& M^{\prime \mu \nu \lambda}_{q-spin} , \\
 M^{\mu \nu \lambda}_{q - OAM}
 &=& \bar{\psi} \,\gamma^\mu \,(\,x^{\nu} \,i \,
 D^{\lambda} 
 \ - \ x^{\lambda} \,i \,
 D^{\nu} \,) \,\psi 
 \ \ \neq \ \  
 M^{\prime \mu \nu \lambda}_{q-OAM} , \\
 M^{\mu \nu \lambda}_{g - spin} 
 &=& M^{\prime \mu \nu \lambda}_{g-spin} , \\
 M^{\mu \nu \lambda}_{g - OAM} 
 &=& M^{\prime \mu \nu \lambda}_{g-OAM}
 \ + \ 2 \,\mbox{Tr} \,[\, (\,D_{\alpha} \,F^{\alpha \mu} \,)
 \,(\,x^{\nu} \,A^{\lambda}_{phys} \ - \ 
 x^{\lambda} \,A^{\nu}_{phys} \,) \,] .
\end{eqnarray}
The difference with the decomposition (II) appears in the
orbital angular momentum parts. 
A great advantage of the decomposition (I) over the decomposition (II)
is the concrete connection with high-energy deep-inelastic-scattering
observables, as we shall argue in the next section.

\section{Observability of our nucleon spin decomposition}

Inserting our decomposition (I) into the helicity normalization condition,
\begin{eqnarray}
 \langle P,s \,|\, W^{\mu} s_{\mu} \,|\, P,s \rangle \,/ \,
 \langle P,s \,|\, P,s \rangle \ = \ \frac{1}{2} ,
\end{eqnarray}
where
\begin{eqnarray}
 W^{\mu} \ = \ - \,\frac{1}{2 \,\sqrt{P^2}} \,\, 
 \epsilon_{\mu \alpha \beta \gamma} \,
 M^{0 \alpha \beta} \,P^{\gamma},
 \ \ \ W^\mu \,s_\mu \ = \ 
 \mbox{\boldmath $J$} \cdot \mbox{\boldmath $P$} \,/ \,
 | \mbox{\boldmath $P$} | ,
\end{eqnarray}
with $W^\mu$ being the standard Pauli-Lubansky vector constructed
from the angular-momentum tensor and the nucleon momentum,
we can derive the following nucleon spin sum rule \cite{Waka10B} :
\begin{eqnarray}
 \frac{1}{2} \ = \ S_q \ + \ L_q \ + \ S_g \ + \ L_g
 \ = \ J_q \ + \ J_g ,
\end{eqnarray}
with
\begin{eqnarray}
 S_q &=& \frac{1}{2} \,\Delta q, \\
 L_q &=& \frac{1}{2} \,\left[ \,
 A_{20}^q (0) \ + \ 
 B_{20}^q (0) \,\right]
 \ - \ \frac{1}{2} \,\Delta q, \\
 S_g &=& \Delta g, \\
 L_g &=& \frac{1}{2} \,\left[\,
 A_{20}^g (0) \ + \ 
 B_{20}^g (0) \,\right]
 \ - \ \Delta g .
\end{eqnarray}
Here, $A^{q/g}_{20} (0)$ and $B^{q/g}_{20}(0)$ respectively stand for the 2nd
moments of the unpolarized GPDs $H^{q/g} (x,\xi,t)$ and $E^{q/g}(x,\xi,t)$ with
$\xi = t = 0$, i.e.
\begin{eqnarray}
 A^{q/g}_{20} (0) &=& \int_{-1}^1 \,x \,H^{q/g} (x,0,0) \,dx ,\\
 B^{q/g}_{20} (0) &=& \int_{-1}^1 \,x \,E^{q/g} (x,0,0) \,dx . 
\end{eqnarray}
As desired, the total nucleon spin consists of four terms,
corresponding to the intrinsic quark spin, the quark OAM, the intrinsic
gluon spin, and the gluon OAM. Our derivation insures that this
decomposition is not only gauge-invariant but also basically
Lorentz frame-independent.

Crucially important to establish is the relation with actual high-energy
observables. We can prove that the quark and gluon intrinsic spin parts
of our recommendable decomposition precisely coincides with the 1st moments
of the polarized distribution functions appearing in the
polarized DIS cross sections \cite{Waka10B}.
\begin{eqnarray}
 \Delta q \ = \ \int_{-1}^1 \,\Delta q(x) \,dx, 
 \ \ \ \ 
 \Delta g \ = \ \int_{-1}^1 \,\Delta g(x) \,dx .
\end{eqnarray}
What is more, we can verify that the following important relation holds : 
\begin{eqnarray}
 L_q &=& J_q \ - \ \frac{1}{2} \,\Delta q \nonumber \\
 &=& \frac{1}{2} \,\int_{-1}^1 \,x \,
 [\,H^q (x,0,0) \ + E^q (x,0,0) \,] \,dx \ - \ 
 \frac{1}{2} \,\int_{-1}^1 \,\Delta q(x) \,dx \nonumber \\
 &=& \langle p \uparrow \,| \,
 M^{012}_{q-OAM} \,| \,
 p \uparrow \rangle ,
\end{eqnarray}
with
\begin{eqnarray}
 M^{012}_{q-OAM} 
 \ = \ \bar{\psi} \,\left(\mbox{\boldmath $x$} \times \frac{1}{i} \,
 \mbox{\boldmath $D$} \right)^3 \,\psi
 &\neq& \left\{ \,
 \begin{array}{l}
 \bar{\psi} \,\left(\mbox{\boldmath $x$} \times \frac{1}{i} \,
 \nabla \right)^3
 \,\psi \\ 
 \bar{\psi} \,\left(\mbox{\boldmath $x$} \times \frac{1}{i} \,
 \mbox{\boldmath $D$}_{pure} \right)^3 \,\psi .
 \end{array}
 \right.
\end{eqnarray}
This identity means that the quark OAM defined as the difference between the
2nd moments of unpolarized GPDs $H + E$ and the 1st moment of polarized
quark distribution just coincides with the proton matrix element of
the our quark OAM operator containing full gauge covariant
derivative \cite{Ji97}. 
This confirms that the quark OAM extracted from the combined analysis of GPDs
and polarized PDFs is the {\it dynamical} OAM not the {\it canonical}-like OAM !

Similarly, we can prove the following identity for the gluon part : 
\begin{eqnarray}
 L_g &=& J_g \ - \ \Delta g \nonumber \\
 &=& \frac{1}{2} \,\int_{-1}^1 \,x \,[\,
 H^g (x,0,0) \ + \ E^g (x,0,0) \,] \,dx \ - \ 
 \int_{-1}^1 \,\Delta g(x) \,dx \nonumber \\
 &=& \langle p \uparrow \,| \,
 M^{012}_{g-OAM} \,| \,
 p \uparrow \rangle ,
\end{eqnarray}
with
\begin{eqnarray}
 M^{012}_{g-OAM} &=& 
 2 \,\mbox{\rm Tr} \,[\,E^j \,(\mbox{\boldmath $x$} \times 
 \mbox{\boldmath $D$}_{pure})^3 \,A^{phys}_j \,] 
 \hspace{10mm} : \ \ \ 
 \mbox{\rm canonical OAM}
 \nonumber \\
 &+& 2 \,\mbox{\rm Tr} \,[\,\rho \,(\mbox{\boldmath $x$} \times
 \mbox{\boldmath $A$}_{phys})^3 \,]
 \hspace{23mm} : \ \ \ 
 \mbox{\rm potential OAM term} .
\end{eqnarray}
It means that the gluon OAM extracted from the combined analysis of GPD and
polarized PDF contains the {\it potential} OAM, in addition to the
{\it canonical-like} OAM. It would be legitimate to call this whole part
the gluon {\it dynamical} OAM.

\section{Some phenomenological implications}

We think it instructive to call attention
to some other recent investigations related to the nucleon spin
decomposition.
As emphasized above, the quark orbital angular momentum
extracted from the combined analysis of the unpolarized GPDs and
the longitudinally polarized quark distribution
functions is the {\it dynamical} orbital angular momentum not the
{\it canonical} one or its nontrivial gauge-invariant extension.
At least so far, we have had no means to extract the {\it canonical}
orbital angular momentum purely experimentally, which also means that
the difference between the {\it dynamical} and {\it canonical}
orbital angular momenta is not a direct experimental observable.
Nevertheless, it is not impossible to estimate the size of this
difference within the framework of a certain model.
In fact, Burkardt and BC estimated the difference between the orbital
angular momentum obtained from the Jaffe-Manohar decomposition
and that obtained from the Ji decomposition within two simple toy models,
and emphasize the possible importance of the vector potential in the
definition of orbital angular momentum \cite{BBC09}.
The difference between the above two orbital angular momenta is nothing
but the {\it potential angular momentum} in our terminology.

Also noteworthy is recent phenomenological investigations on the role
of orbital angular momenta in the nucleon spin. In a recent paper,
we have pointed out possible existence of significant discrepancy
between the lattice QCD predictions \cite{LHPC08},\cite{LHPC10} 
for $L^u - L^d$ (the difference
of the orbital angular momenta carried by up- and down-quarks in the
proton) and the prediction of a typical low energy model of the nucleon,
for example, the refined cloudy-bag model \cite{MT88}.
It is an open question whether
this discrepancy can be resolved by strongly scale-dependent nature
of the quantity $L^u - L^d$ especially in the low $Q^2$
domain  as claimed in \cite{Thomas08}, or whether the discrepancy has
a root (at least partially) in the existence of two kinds of quark orbital
angular momenta as indicated in \cite{Waka10E},\cite{WT05}.
(See also \cite{WN06},\cite{WN08}.)

\section{Summary and conclusion}

To sum up, inspired by the recent proposal by Chen et al., we find it possible
to make a gauge-invariant decomposition of covariant angular-momentum
tensor of QCD in an arbitrary Lorentz frame.
Based on this fact, we could show that our decompositions of nucleon spin are
not only gauge-invariant but also practically frame-independent.
We have also succeeded to convince that each piece of our nucleon spin
decomposition (I) precisely corresponds to the observables that can be extracted
from combined analysis of the GPD measurements and the polarized DIS
measurement, thereby supporting the standardly-accepted program aiming
at complete decomposition of the
nucleon \cite{HERMES06A}\nocite{HERMES06B}-\cite{JLabHallA07}.

A practically very important lesson learned from our theoretical consideration
is that the quark OAM extracted from the combined analysis of GPDs and
polarized PDFs is  the {\it dynamical} quark OAM, not the {\it canonical} OAM or
its non-trivial ``gauge-invariant extension'' as advocated by Chen et al.
Similarly, the gluon OAM extracted from the combined analysis of the gluon GPD
and polarized gluon distribution is the {\it dynamical} gluon OAM, which
contains the {\it potential angular momentum} term in addition to the
{\it canonical} one. 
Note that, at the moment, we do not know any practical means, with which we
can extract the canonical OAMs purely experimentally, i.e. model independently.
Still, one should keep in mind the existence of 2 kinds of quark and gluon OAMs !
 
\section*{Acknowledgment}
This work is supported in part by a Grant-in-Aid for
Scientific Research for Ministry of Education, Culture, Sports, Science
and Technology, Japan (No.~C-21540268)

\end{document}